\documentclass[iop]{emulateapj}
\usepackage{graphicx}
\usepackage{amsmath}
\usepackage[normalem]{ulem}
\usepackage{multirow}
\usepackage{color}
\DeclareMathSizes{14}{14}{14}{14}

\def\be{\begin{equation}}
\def\ee{\end{equation}}

\def\bea{\begin{eqnarray}}
\def\eea{\end{eqnarray}}

\begin{document}

\title[Prompt GRB spectra from Comptonization]{Gamma-ray
  burst spectra and spectral correlations from sub-photospheric
  Comptonization}

\author{Atul Chhotray and Davide Lazzati} \affil{Department of
  Physics, Oregon State University, 301 Weniger Hall, Corvalis, OR
  97331, USA}

\begin{abstract}
  One of the most important unresolved issues in gamma-ray burst
  physics is the origin of the prompt gamma-ray spectrum. Its general
  non-thermal character and the softness in the X-ray band remain
  unexplained. We tackle these issues by performing Monte Carlo
  simulations of radiation-matter interactions in a scattering
  dominated photon-lepton plasma. The plasma -- initially in
  equilibrium -- is driven to non-equilibrium conditions by a sudden
  energy injection in the lepton population, mimicking the effect of a
  shock wave or the dissipation of magnetic energy. Equilibrium
  restoration occurs due to energy exchange between the photons and
  leptons. While the initial and final equilibrium spectra are
  thermal, the transitional photon spectra are characterized by
  non-thermal features such as power-law tails, high energy bumps, and
  multiple components. Such non-thermal features are observed at
  infinity if the dissipation occurs at small to moderate optical
  depths, and the spectrum is released before thermalization is
  complete.  We model the synthetic spectra with a Band function and
  show that the resulting spectral parameters are similar to
    observations for a frequency range of 2-3 orders of magnitude
    around the peak. In addition, our model predicts correlations
  between the low-frequency photon index and the peak frequency as
  well as between the low- and high-frequency indices. We explore
  baryon and pair dominated fireballs and reach the conclusion that
  baryonic fireballs are a better model for explaining the observed
  features of gamma-ray burst spectra.
\end{abstract}


\section{Introduction}
The radiation mechanism that produces the bulk of the prompt emission
of Gamma-Ray Bursts (GRBs) is still a matter of open debate
(e.g. Mastichiadis \& Kazanas 2009; Medvedev et al. 2009; Ryde \&
Pe'er 2009; Asano et al. 2010; Ghisellini 2010; Lazzati \& Begelman
2010; Daigne et al. 2011; Massaro \& Grindlay 2011; Resmi \& Zhang
2012; Hasco\"et et al. 2013; Crumley \& Kumar 2013).  Among the many
proposed possibilities, the synchrotron shock model (SSM) and the
photospheric model (PhM) have recently gathered most of the attention
(Rees \& Meszaros 1994; Piran 1999; Lloyd \& Petrosian 2000;
M{\'e}sz{\'a}ros \& Rees 2000; Rees \& M{\'e}sz{\'a}ros 2005; Giannios
2006; Pe'er et al. 2006; Bo{\v s}njak et al. 2009; Lazzati et
al. 2009; Beloborodov 2010; Mizuta et al. 2011; Nagakura et
al. 2011). Within the SSM, the bulk of the prompt radiation is
produced by synchrotron from a non-thermal population of electrons
gyrating around a strong, locally-generated magnetic field. The
non-thermal leptons are produced either by trans-relativistic internal
shocks (the SSM proper, Rees \& Meszaros 1994) or by magnetic
reconnection in a Poynting flux dominated outflow (e.g. the ICMART
model, Zhang \& Yan 2011). The SSM naturally accounts for the broad,
non-thermal nature of the spectrum. However, it has difficulties in
accounting for bursts with particularly steep low-frequency slopes
(Preece et al. 1998; Ghisellini et al. 2000) and has limited
predictive power, since the radiation properties are tied to poorly
constrained quantities such as the lepton's energy distribution, the
ad-hoc equipartition parameters, and the ejection history of shells
from the central engine.

The PhM does not specify a radiation mechanism, assuming instead that
the burst radiation is produced in the optically thick part of the
outflow and advected out, its spectrum being the result of the strain
between mechanisms that tend to bring radiation and plasma in thermal
equilibrium and mechanisms that can bring them out of balance (e.g.,
Beloborodov 2013). The PhM has been shown to be able to reproduce
ensemble properties of the GRB population, such as the debated Amati
correlation, the Golenetskii correlation, and the recently discovered
correlation between the burst energetics and the Lorentz factor of the
outflow (Amati et al. 2002; Amati 2006; Liang et al. 2010; Fan et
al. 2012; Ghirlanda et al. 2012; Lazzati et al. 2013; L\'opez-C\'amara
et al. 2014). However, it is not yet understood how the broad-band
nature of the prompt spectrum, spanning many orders of magnitude in
frequency, is produced.  In a hot, dissipationless flow, only the
adiabatic cooling of the plasma would work as a mechanism to break
equilibrium, and the GRB outflow would work as a miniature big bang,
the entrained radiation maintaining a Planck spectrum. In a cold,
dissipationless outflow, lepton scattering dominates the
radiation-matter interaction producing a Wien spectrum (Rybicki \&
Lightman 1979). Outflows from GRB progenitors are, however, far from
dissipationless. Hydrodynamic outflows are continuously shocked out to
large radii (Lazzati et al. 2009), and Poynting-dominated outflows
suffer dissipation through magnetic reconnection (Giannios \& Spruit
2006). Either way, even if thermal equilibrium is reached at some
point in the outflow, it is likely that such equilibrium is broken by
a sudden release of energy in the lepton population or altered by a
slow and continuous (or episodic) injection of energy. The effects of
such energy injection on the photospheric spectrum are profound (e.g.,
Giannios 2006; Pe'er et al. 2006; Beloborodov 2010; Lazzati \& 
Begelman 2010). In addition, the interaction between different parts
of the outflow in a stratified flow alter the thermal spectra into a
non-thermal, highly polarized spectrum (Ito et al. 2013, 2014; Lundman
et al. 2013).

In this paper we investigate the evolution of the radiation spectrum
following the sudden injection of energy in the lepton population of a
plasma, assuming that the radiation and leptons interact via Compton
scattering and pair processes. We use a Monte Carlo (MC) method that
evolves simultaneously the photon and lepton populations by performing 
inelastic scattering between photons and leptons in both the
non-relativistic and the relativistic (Klein-Nishina) regimes. The
code also accounts for $e^-e^+$ annihilation (pair annihilation
henceforth) and $e^-e^+$ pair production from photon-photon collisions
(pair production henceforth).  We focus on transient features that can
be observed if the episode(s) of energy injection in the leptons occur
at small or moderate optical depths ($\tau<1000$).

This manuscript is organized as follows. In Section 2 we describe the
physics and the methods of the MC code, in Section 3 we show our
results and in Section 4 we discuss the results and compare them to
previous findings.

\section{Methodology}

\subsection{Step 1: Particle Generation}
As a first task, the code generates a user-defined number of leptons
and photons. Their energies follow a distribution that can be either
of thermal equilibrium (Wien for the photons and Maxwell-J\"uttner for
the leptons) or any other user-specified distribution. After
initializing the photon and lepton distributions, our code performs
the following steps iteratively.

\subsection{Step 2: Particle/Process selection}
To initiate either a scattering or a pair event we need to select two
particles\footnote{Note that here particle can mean both a lepton or a
  photon.} - which we obtain by randomly selecting a pair from our
generated distributions.  Depending upon the particles selected,
Compton scattering (if a photon and a lepton is chosen), pair
annihilation (if an $e^-$ or $e^+$ is chosen) or pair production (if
two photons are chosen) is performed or another pair is re-selected if
any other combination occurs. After the selection, the code proceeds
with the following calculations:
\begin{itemize}
\item[1.] Incident angle generation ($\theta$) using the appropriate
  relativistic scattering rates, under the assumption that both
  leptons and photons are isotropically distributed.
\item[2.] Lorentz boost to the necessary reference frames (details
  explained in successive sections) from the lab frame.
\item[3.] Event probability computations from total cross section
  ($\sigma$) calculations.
\item[4.] Scattering angle generation from differential cross section
  ($\frac{d\sigma}{d\Omega}$).
\item[5.] Lorentz boost from the necessary frame back to the lab
  frame.
\end{itemize}
In the following sub-sections we discuss each of the three possible
processes in detail

\subsubsection{Process 1: Compton Scattering}
As the choice of reference frame is arbitrary, in the lab frame we can
assume that the lepton is traveling along the x-axis and the photon is
incident upon the lepton in the xy plane without any loss of
generality. The angle of incidence $\theta_{\gamma e}$ between the
chosen photon-lepton pair is generated by a probability distribution
$P_{\gamma e}$:
\be 
P_{\gamma e}(\beta_{e}, \theta_{\gamma e}) \propto \sin\theta_{\gamma e} (1 -
\beta_{e} \cos\theta_{\gamma e}) \label{eq:prob_cs} 
\ee

\noindent
where $\beta_{e}=v_e/c$, is the ratio of lepton speed to the speed of
light.\\*
To simulate the scattering event  
the code Lorentz transforms to the lepton frame (that we call the 
co-moving frame). The probability that the chosen photon-lepton 
pair interacts depends
on the incident photon energy in the co-moving frame. As Compton
scattering becomes less efficient at higher energies, photons having
energies comparable to or greater than the lepton's rest mass energy
are less likely to scatter.  Using Monte Carlo sampling we determine
if scattering occurs or not. This is done by generating a random
number and comparing it to the ratio of the Klein-Nishina cross
section $\sigma_{\gamma e}$ to the Thomson cross section, which we use
as a reference value. We proceed with the scattering event if
$\sigma_{\gamma e}/\sigma_T \geq s_1$ where $s_1$ is a random
number. If the condition is not satisfied, the code returns to step
2. If instead the condition is satisfied and the scattering occurs, 
the code generates the polar scattering angle $\theta'_s$
in accordance with the Klein-Nishina differential cross-section
formula
\be 
\frac{d\sigma_{\gamma e}}{d\Omega} =
\frac{r^2_0}{2}\frac{E_s'^2}{E'^2}\left(\frac{E'}{E'_s}
  + \frac{E'_s}{E'} - \sin^2 \theta'_s \right) \label{eq:KN}
\ee 
where $r_0$ is the classical radius of an electron, 
$E'$ and $E'_s$ are the energies of the incident and scattered 
photon respectively (e.g. Bluementhal \& Gould 1970, Longair 2003 and 
Rybicki \& Lightman 1979).  The energy transfer equation connecting
$E'$ with $E'_s$ is the Compton equation (e.g. Bluementhal \& Gould
1970, Longair 2003 and Rybicki \& Lightman 1979)
\be E'_s = \frac{E'}{1 + \frac{E'}{m_e c^2}(1 -
  \cos\theta'_s)}.\label{eq:etransfer} 
\ee 
(Note here that $\theta'_s$ is the angle that the scattered photon
makes with the direction of propagation of the incident photon in the
co-moving frame. Hence equations~(\ref{eq:KN}) and~(\ref{eq:etransfer}) 
hold true only in the lepton frame). Finally, the azimuthal angle 
$\phi'_s$ is generated randomly between zero and $2\pi$.  Thus, we 
now have the four momenta of the scattered particles in the 
co-moving frame.

\subsubsection{Process 2: Pair Production / Photon Annihilation}
If the particle selection process selects two photons then the pair
production/photon annihilation channel is chosen. The code 
computes the angle of incidence $\theta_{\gamma \gamma}$ between the
chosen photons by using the probability distribution $P_{\gamma
  \gamma}$:
\be
P_{\gamma \gamma}(\theta_{\gamma \gamma}) \propto \sin{\theta_{\gamma \gamma}} (1 - \cos{\theta_{\gamma \gamma}}).
\ee
To ensure that the photon pair has enough energy to lead to a pair 
production event the code checks the energy of the photon/s 
in the zero momentum frame. The 
zero momentum frame photon energy $E'_o$ can be computed
given the incident photon energies 
$E_1$, $E_2$ and the incident angle as 
\be
E'_o = \sqrt{E_1 E_2} \sin(\theta_{\gamma \gamma}/2).
\ee
(Gould \& Schereder 1967). If $E'_o < m_e c^2$ the colliding photon 
pair is not energetic energy to produce an $e^- e^+$ pair, 
hence the code jumps to step 2 for a new particle pair selection. 
Due to the energy dependence of cross-section $\sigma_{\gamma\gamma}$,  
even photons exceeding the energy threshold might not produce 
pairs. To make this determination, we again use the 
Thomson cross section as a reference and determine if the 
photon annihilation takes place by randomly drawing one 
number $s_2$, obtaining $\sigma_{\gamma\gamma}$ by boosting 
to the center of momentum frame and evaluating if 
$\sigma_{\gamma\gamma}/\sigma_T \geq s_2$. If the 
inequality holds true, the code proceeds with the pair 
production calculation. Otherwise, it is abandoned and 
the code returns to step 2.\\*
Once the photons succeed in producing leptons, the polar scattering
angle $\theta'_s$ of the newly born $e^-$ is computed from the pair 
annihilation differential cross section as given by
\be
\frac{d\sigma_{{\gamma \gamma}}}{d\Omega} = \frac{r_0^2 \pi}{2}b\left(\frac{m_e c^2}{E'_o}\right)^2 \frac{1 - b^4 \cos^4{\theta'_s} + 2\left(\frac{m_e c^2}{E'_o}\right)^2 b^2 \sin^2{\theta'_s}}{\left(1 - b^2 \cos^2{\theta'_s}\right)^2}. \label{eq:dcs_pp}
\ee
(see Jauch \& Rohrlich 1980, p.300) where 
$b=\sqrt{1 - \left(\frac{m_e c^2}{E'_o}\right)^2}$. 
A random azimuthal angle $\phi'_s
\in [0,2 \pi)$ is assigned to the $e^-$. Note that Lorentz
transformation to the zero momentum frame is necessary because 
equation~(\ref{eq:dcs_pp}) is frame dependent. Utilizing
conservation laws, the four momenta of the $e^+$ can be determined.

\subsubsection{Process 3: Pair Annihilation/ Photon Production}

The pair annihilation channel is chosen if the random particle
selection constitutes an $e^-e^+$ pair. As with the other channels, we
first determine the incident angle $\theta_{ee}$ (subscript $_{ee}$
stands for lepton pair annihilation) between the pair
by computing the probability distribution of scattering as given by
\be
P_{ee}(\beta_{e^-},\beta_{e^+},\theta_{ee}) \propto \sin{\theta_{ee}} f_{kin} \label{eq:prob_pa}
\ee
where $f_{kin}$ as obtained from (Coppi \& Blandford 1990) is 
given by:
\be
f_{kin}=\sqrt{\beta_{e^-}^2 + \beta_{e^+}^2 - \beta_{e^-}^2 \beta_{e^+}^2 \sin^2{\theta_{ee}} - 2 \beta_{e^-} \beta_{e^+} \cos{\theta_{ee}}}. 
\ee
Here $\beta_{e}=\frac{v_e}{c}$ i.e. the ratio of lepton speed to the
speed of light. The code transforms all quantities to the rest frame of
the electron to calculate the the total cross section $\sigma_{ee}$ as
(Jauch \& Rohrlich 1980, p.269):
\be
\sigma_{ee}=\frac{r_0^2 \pi}{\beta'^2} \left[ \frac{\left( \gamma' + \frac{1}{\gamma'} + 4 \right) \ln{(\gamma' + \sqrt{\gamma'^2 - 1}}) - \beta'(\gamma' + 3)}{\gamma'(\gamma' + 1)}\right] \label{eq:tcs_pa}
\ee
where $\beta'=v'_{e+}/c$, $\gamma' = \frac{1}{\sqrt{1 - \beta'^2}}$
i.e.  the $e^+$ speed and Lorentz factor respectively in the co-moving
frame traveling with the $e^-$. On comparing the
$\sigma_{ee}/\sigma_T$ with a random number $s_3$ the code
evaluates the occurrence of the annihilation event.  If the event
fails, the code returns to step 2 to re-select another pair of
particles. Following a successful event, the polar scattering angle
$\theta'_s$ between either of the pair produced photons is generated
from the differential cross section (from Jauch \& Rohrlich 1980,
p.268)
\be
\frac{d\sigma_{ee}}{d\Omega}= \frac{r_0^2 \pi}{\beta' \gamma' d} \left[ \gamma' + 3 - \frac{\left[
1+d \right]^2 }{(1 + \gamma')d} - \frac{2 (1 + \gamma')d}{\left[ 1 + d \right]^2} \right] \label{eq:dcs_pa}
\ee
where $x=\cos{\theta'_s}$ and $d=\gamma'(1 - \beta' x)$. As pointed 
out in the preceding sub-sections, Lorentz transformation to the 
electron frame is necessary as equation~(\ref{eq:dcs_pa}) is expressed 
in terms of quantities defined in the electron's co-moving frame. 
The random azimuthal angle $\phi'_s \in [0,2 \pi)$ is randomly 
assigned to either photon. Using conservation laws, the four 
momenta of the pair produced photons can be obtained.

\subsection{Step 3: Back to the lab frame}
At the end of the event, the code transforms the
four momenta of the particles back to the lab frame by employing
Lorentz transformations. The loop is repeated until equilibrium is
restored i.e. when the particle numbers saturate 
and distributions become thermal.

\section{Results}
We employ the Monte Carlo code described above to study the evolution
of the radiation spectrum in a closed box containing leptons and
photons. The simulations are initialized with a Wien radiation
spectrum at $10^6$~K and a non-equilibrium lepton population, either
because leptons and photons are at different temperature or because
the leptons energy distribution is non-thermal.  This is expected to
mimic a scenario in which the leptons and radiation were initially at
equilibrium, but the lepton population has been brought out of
equilibrium by a sudden energy release. Such energy release may be due
to shocks in the fluid (e.g., Rees \& Meszaros 1994; Lazzati \&
Begelman 2010) or by magnetic reconnection in a magnetized outflow
(e.g. Giannios \& Spruit 2006, McKinney \& Uzdensky 2012). As it will
be clear at the end, a fundamental parameter that determines the
interaction between the photons and leptons is the particle ratio,
i.e., the ratio of photon and lepton number densities. In a GRB
outflow, such a ratio can be readily estimated.

Let us call $E_K$ the kinetic energy of the outflow carried by
particles with non-zero rest mass and $E_\gamma$ the
energy in electromagnetic radiation. We have:
\begin{eqnarray}
\frac{E_\gamma}{E_K}&=&\frac
{N_\gamma h\nu_{\rm{pk}}}
{\left(N_p+\frac{m_e}{m_p}N_{\rm{lep}}\right)\Gamma m_p c^2} \nonumber
\\ 
&\simeq&
10^{-5}\frac{n_\gamma}{n_p+\frac{n_{\rm{lep}}}{1836}}
\left(\frac{h\nu_{\rm{pk}}}{1\,{\rm{MeV}}}\right)
\Gamma_2^{-1}
\label{eq:partratio}
\end{eqnarray}
By calling $\eta=E_\gamma/(E_\gamma+E_K)$ the radiative efficiency of
the outflow, and assuming that matter and radiation are coupled in the
optically thick region and occupy the same volume,
equation~(\ref{eq:partratio}) can be inverted to yield:
\begin{equation}
\frac{n_\gamma}{n_{\rm{lep}}}=\left\{
\begin{array}{ll}
10^5\frac{\eta}{1-\eta}\left(\frac{1\,{\rm{MeV}}}{h\nu_{pk}}\right)\Gamma_2
& \qquad n_{\rm{lep}}=n_p\\
50\frac{\eta}{1-\eta}\left(\frac{1\,{\rm{MeV}}}{h\nu_{pk}}\right)\Gamma_2
& \qquad n_{\rm{lep}}\gg n_p
\end{array}
\right.
\label{eq:partratio2}
\end{equation}
where the top line is valid for a non-pair enriched fireball while the
bottom line is for a pair-dominated fireball. All values in between
are allowed for a partially pair-enriched fireball. Note also that we
used the convention $\Gamma_2=\Gamma/10^2$. GRB fireballs are
therefore photon-dominated, even if highly pair-enriched.

We here consider two possible values of the particle ratio. As a 
representative of pair-enriched plasma, we explore the case
$n_\gamma/n_{\rm{lep}}=10$. A non-enriched plasma 
(or photon-rich plasma) is represented by the ratio 
$n_\gamma/n_{\rm{lep}}=1000$. Note that the latter value is not as
extreme as the one in equation~(\ref{eq:partratio2}). It is, however,
technically challenging to simulate any higher value of the particle
ratio. To ensure that the statistics of the lepton population is under
control, we need to simulate at least 1000 irreducible electrons
(electrons that are not possibly annihilated by a positron). For a
particle ratio $n_\gamma/n_{\rm{lep}}=10^5$, that would require the
simulation of $10^8$ photons.  We believe that the adopted value
$n_\gamma/n_{\rm{lep}}=1000$ does capture the characteristics of the
spectrum emerging from a photon-rich plasma and we will discuss the
consequences of higher particle ratios in Section~4.

For each particle ratio, we explore different scenarios in which the
accelerated leptons are either thermal (Lazzati \& Begelman 2010) or
non-thermal (e.g. Giannios 2006; Pe'er et al. 2006; Beloborodov 2010)
and we consider the possibility of multiple acceleration events, in
which the leptons are re-energized before the equilibrium is
reached. Some of these possibilities have been previously explored, in
particular the Comptonization from a non-thermal population of
electrons (e.g. Pe'er et al. 2006). We do not consider in this study
continuous energy injection, in which a stationary equilibrium between
photons and electrons is reached, and for which our code is not well
suited (e.g. Giannios 2006; Pe'er et al. 2006).

All simulations are run until equilibrium is attained. Here we define
equilibrium as the time at which the spectral shape does not change
with further collisions and the number of photons and leptons
saturate.  This is generally much later than the time at which the
total energies in leptons and photons approach their asymptotic
values, since a very small amount of energy can make a significant
difference in the tails of the distribution, which are the interesting
aspect of the spectrum for this study.  Our simulations do not have a
time stamp, since all processes involved are scale free. A time stamp
can be added upon deciding on a particle and photon density, rather
than a total number as specified in the code. A meaningful comparison
with the data can be accomplished by considering that a photon in a
relativistic outflow with Thomson opacity $\tau$ scatters-off/collides
with leptons an average number of times $n_{\rm{sc}}\simeq\tau$ before
being detected by an observer at infinity (e.g. Pe'er et al. 2005).
Here we adopt as the Thomson opacity of a medium $\tau = \int n_{lep}
\sigma_T ds$ (see Rybicki \& Lightman 1979).  It is possible
therefore to look at our spectra in the following manner: if a shock
or a reconnection event dissipates energy in the outflow at a certain
optical depth $\tau$, the spectrum observed at infinity is the one
derived from our code after $\tau_{diss}$ scatterings per
photon.

\subsection{Photon Rich Plasma}
We first explore a photon-rich plasma with 
$n_\gamma/n_{\rm{lep}}=1000$. Three simulations are initialized with an
out-of-equilibrium electron population (there are no positrons
initially in the plasma) with different initial 
distributions. We inject identical amounts of total 
kinetic energy $K$ in all three cases, raising the average 
kinetic energy of the leptons to 1.365~MeV. This can be 
considered as a mild energy injection and within the 
equipartition shock-acceleration scenario this corresponds to either a
mildly-relativistic shock or a relativistic shock with a fairly low
fraction of energy given to electrons ($\epsilon_e\ll1$) 
(e.g. Guetta et al 2001). In the first simulation, the leptons adopt 
a Maxwell-J\"uttner distribution at $T_e=6.5\times10^{9}$~K. In 
the second case the
leptons conform to a Maxwellian distribution (at $10^8$~K)
which is smoothly connected to a non-thermal power law tail
$n_e(\gamma)\propto\gamma^{-2.2}$. The third and final simulation 
explores the scenario of energy dissipation via multiple (10) 
less energetic injections instead of a single intense injection 
event.

\subsubsection{Thermal Leptons at $6.5 \times 10^9$~K}
\label{sect:thermal_phrich}
The lepton population in this case is shocked and then 
thermalizes at $6.5 \times 10^{9}$ K. A similar scenario was 
explored analytically and with a simplified Monte Carlo code by 
Lazzati \& Begelman (2010).

\begin{figure}
\includegraphics[width=80mm]{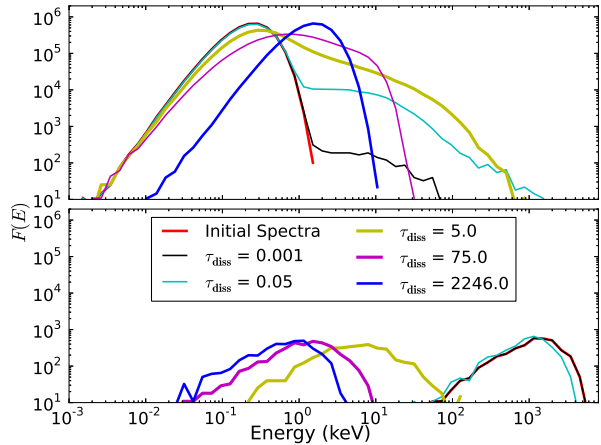}
\vskip 4pt
\caption{Radiation spectrum (upper panel) and leptons' kinetic energy
  distribution (lower panel) at different simulation stages for a
  photon-rich plasma ($N_\gamma/N_{\rm{lep}}=1000$) with a sudden
  injection of thermal energy in the lepton population (see
  Sect.~\ref{sect:thermal_phrich}). The legend displays the various 
  optical depths at which if energy was injected, the 
  corresponding color coded spectrum and distribution would be 
  observed.}
\label{fig:MJ_6pt5e9_3_W_6_6}
\end{figure}
\begin{figure}
\includegraphics[width=80mm]{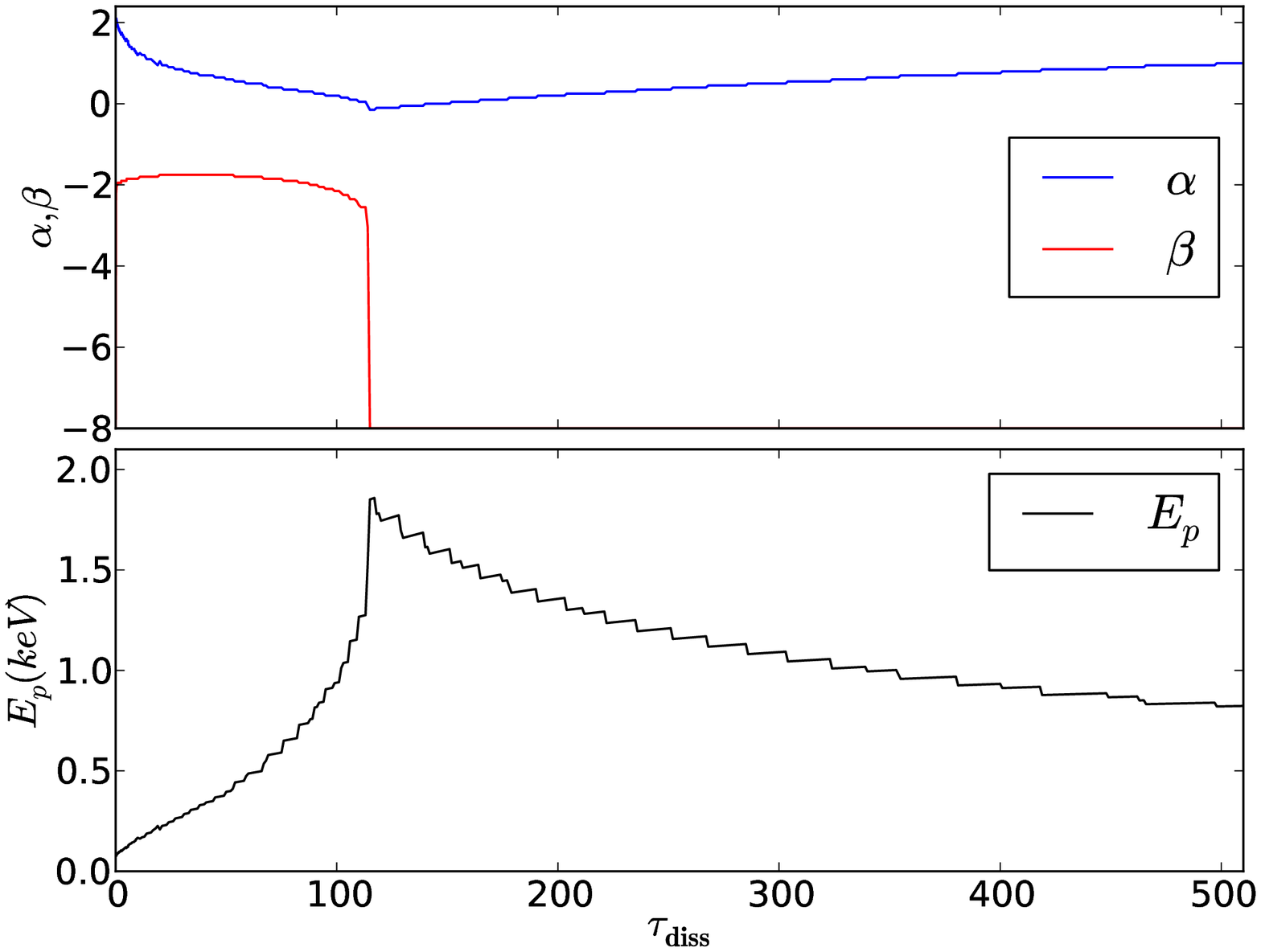}
\vskip 4pt
\caption{Evolution of the Band parameters $\alpha$, $\beta$ and, $E_{p}$ of
  spectra from the simulation shown in Figure~\ref{fig:MJ_6pt5e9_3_W_6_6}.
  The x-axis indicates the optical depth of energy injection.}
\label{fig:ABEPLOT_MJ_6pt5e9_3_W_6_6}
\end{figure}
\begin{figure}
\includegraphics[width=80mm]{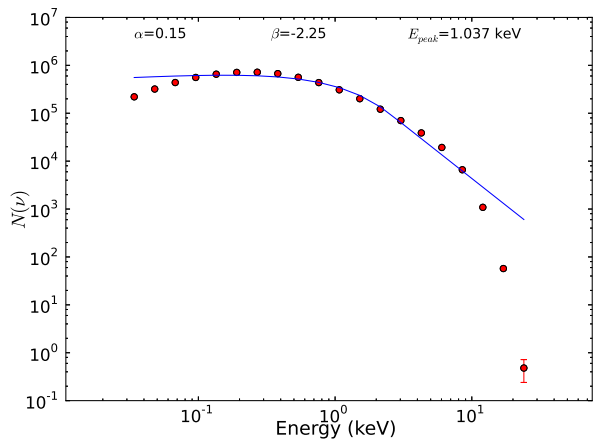}
\vskip 4pt
\caption{Fitting of the Band parameters $\alpha$, $\beta$ and $E_{p}$ of
  spectra from the simulation shown in Figure~\ref{fig:MJ_6pt5e9_3_W_6_6}
  at $\tau_{diss}=103$.}
\label{fig:BANDS_MJ_6pt5e9_3_W_6_6_103e6}
\end{figure}
The results of the simulation are shown in
Figure~\ref{fig:MJ_6pt5e9_3_W_6_6}, where the evolution of the
radiation spectrum and of the spectrum of the kinetic energy of the
leptons' population are displayed. We first note that the final
distributions (blue curves in both panels) are all thermal, as
expected for a plasma in equilibrium. Looking at the intermediate
spectra in more detail, we notice that the immediate reaction of the
radiation spectrum is the formation of a high-frequency non-thermal
tail, initially appearing as a new component (for spectra at
$\tau_{diss}=0.001$) and subsequently forming a continuous tail
stemming from the thermal photon population ($\tau_{diss}=5$). At a
subsequent stage, the low-frequency part of the radiation spectrum is
also modified, with the spectral peak migrating to higher frequencies
and causing a flattening of the low-frequency component
($\tau_{diss}=75$). The figure shows that the spectrum takes a very
large number of scatterings for equilibrium restoration, especially
for frequencies lower than the peak. For energy dissipation at optical
depths up to $\sim100$ a high-frequency non-thermal tail is
observed. A non thermal low-frequency tail is instead observed even
for a larger optical depth, up to a few thousand.

In order to quantify our synthetic transient spectra and compare them
with observations, we fit them to an analytic model. We adopt the
widely used Band function (Band et al. 1993) and fit it to the data
over a frequency range of three orders of magnitude. Although the GRB
spectra are in most cases more complex than a Band function
(e.g. Burgess et al. 2014, Guiriec et al. 2011, 2013) this still
constitutes a zero-order test that any model should pass. We begin by
computing the mean frequency from our data and select neighboring
frequencies within 1.5 orders of magnitude around the mean. This data
set is binned in frequency and a best-fit Band function is obtained by
minimizing the $\chi^2$.

Figure~\ref{fig:ABEPLOT_MJ_6pt5e9_3_W_6_6} shows the evolution of the
spectral parameters $\alpha$, $\beta$ and $E_p$ for increasing optical
depths. We again emphasize that this should not be considered as a
time evolution, since the number of scattering is set by the optical
depth at which the energy is released in the leptons. A sample fit of
the spectrum at $\tau_{diss}=103$ to the Band function is shown in
Figure~\ref{fig:BANDS_MJ_6pt5e9_3_W_6_6_103e6}. The figure
  represents a typical case, and shows that the Band model fits well
  the frequencies around the peak but deviations are observed for the
  lowest and highest frequencies. We will address this issue further
  in the discussion. The legend at the top of the figure shows the
Band parameters for the fit. An interesting aspect of these
simulations is that the low-frequency photon index $\alpha$ and the
peak frequency are strongly anti-correlated. This is due to the fact
that it is necessary that the peak frequency shifts to higher values
for the low-frequency spectrum to change from its thermal equilibrium
shape. We also note that the high-energy slope anticipates the
low-energy one, the non-thermal features building-up earlier and
disappearing faster. We will discuss in more detail these correlations
and their implications in Section~4.

\subsubsection{Maxwellian leptons at $10^8$ K with a power law tail $p=2.2$}
\label{sect:nonthermal_phrich}
\begin{figure}
\includegraphics[width=80mm]{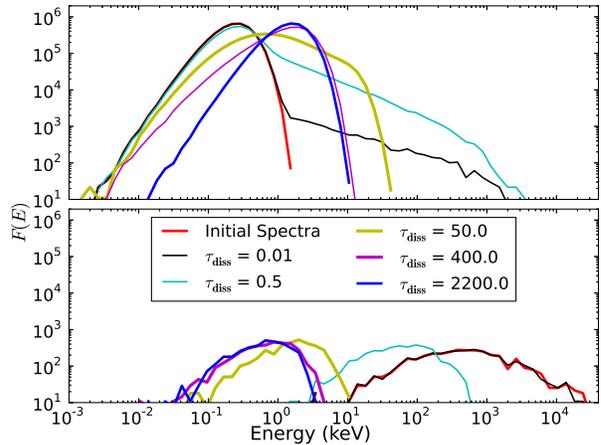}
\vskip 4pt
\caption{Color coded photon spectrum (upper panel) and leptons' kinetic energy
  distribution (lower panel) at different stages for the photon-rich simulation 
  discussed in Section~\ref{sect:nonthermal_phrich}. The legend displays the 
  various optical depths at which if energy was injected, the 
  corresponding color coded spectrum and distribution would be observed.}
\label{fig:MJNT_8-2.2_3_W_6_6}
\end{figure}
\begin{figure}
\includegraphics[width=80mm]{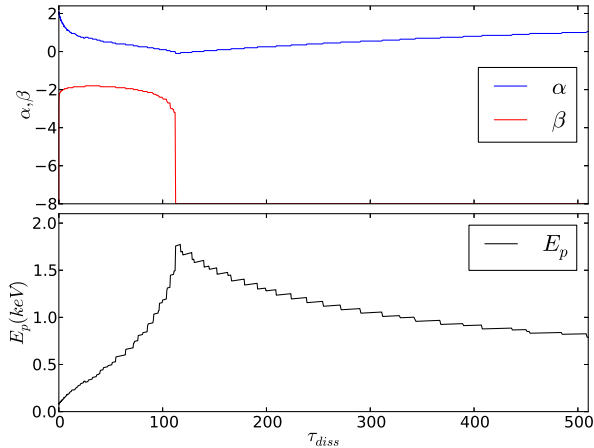}
\vskip 4pt
\caption{Evolution of the Band parameters $\alpha$, $\beta$ and
  $E_{p}$ of spectra from the simulation shown in
  Figure~\ref{fig:MJNT_8-2.2_3_W_6_6}. The x-axis indicates the
  optical depth of energy injection.}
\label{fig:ABEPLOT_MJNT_8-2pt2_3_W_6_6}
\end{figure}
Most models of internal shocks predict the acceleration of non-thermal
particles. Comptonization of seed thermal photons by non-thermal
leptons has been widely studied in different scenarios and under
different assumptions (e.g. Giannios 2006; Pe'er et al. 2005,
2006). In this scenario the shock generates a non-thermal lepton
distribution characterized by \be N(E)dE \propto \gamma^{-p}
d\gamma \label{eq:ntdistr} \ee where $\gamma$ is the lepton Lorentz
factor and $p=2.2$.  The results of the simulation are shown in
Figure~\ref{fig:MJNT_8-2.2_3_W_6_6}, where we present the evolving
radiation spectrum and distribution of the kinetic energy of the
leptons' population. We notice that the equilibrium photon and lepton
distributions (blue curves) are thermal, as expected at
equilibrium. We also notice that the spectrum appears non-thermal for
a wide range of opacities. Initially a prominent high-energy power-law
tail is developed, for a very small opacity (or $\tau_{diss} \sim
0.01$).  As the injection opacity increases, the power-law tail is
truncated at progressively lower frequencies, the peak frequency
shifts to higher values, and a non-thermal tail at low-frequencies
develops. The high-frequency tail disappears for $\tau_{diss} \sim
400$, but even larger opacities are required to turn the low-frequency
tail back to the scattering-dominated equilibrium spectrum. We
  fit the Band function to our synthetic spectra and obtain
  Figure~\ref{fig:ABEPLOT_MJNT_8-2pt2_3_W_6_6}, which shows the
  evolution of the spectral parameters $\alpha$, $\beta$ and $E_p$ for
  increasing injection optical depths. We also notice correlations
between the spectral parameters $\alpha$ and the peak frequency, as
discussed in Section~\ref{sect:thermal_phrich}.

\subsubsection{Discrete Multiple Energy Injections}
\label{sect:multipleinj_phrich}
The presence of multiple minor shocks has been emphasized in 2D
axisymmetric numerical simulations of jets in collapsars (e.g. Lazzati
et al. 2009) and seem to be an even more common feature in 3D
simulations (L\'opez-C\'amara et al. 2013). Hence, to provide a more
realistic scenario for the energy injection we explore lepton heating
by multiple energy injections mimicking multiple shocks instead of a
single more powerful one. The total energy injected into the lepton
population is identical to the amount injected in the simulations
discussed in Sections~\ref{sect:thermal_phrich} and
~\ref{sect:nonthermal_phrich}. However, the energy
is divided into 10 equal and discrete partitions with each one being
injected and distributed uniformly among the leptons, after every
million scatterings.
\begin{figure}
\includegraphics[width=80mm]{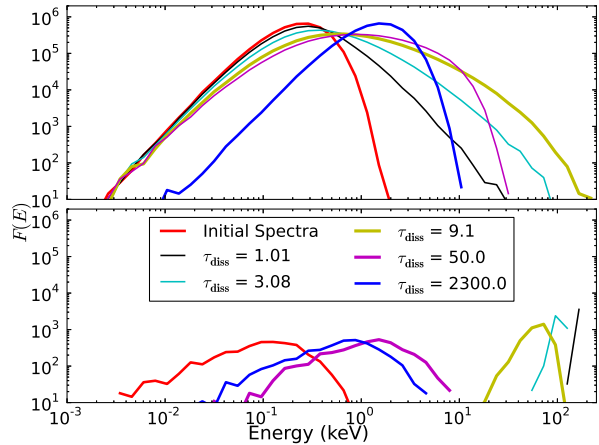}
\vskip 4pt
\caption{Photon spectrum (upper panel) and leptons' kinetic energy
  distribution (lower panel) at different stages of the simulation 
  discussed in Section~\ref{sect:multipleinj_phrich}.
  The legend associates the various optical depths of energy injection with the  
  corresponding color coded spectrum and distribution observed.}
\label{fig:INJ_MB_6_3_W_6_6}
\end{figure}

\begin{figure}
\includegraphics[width=80mm]{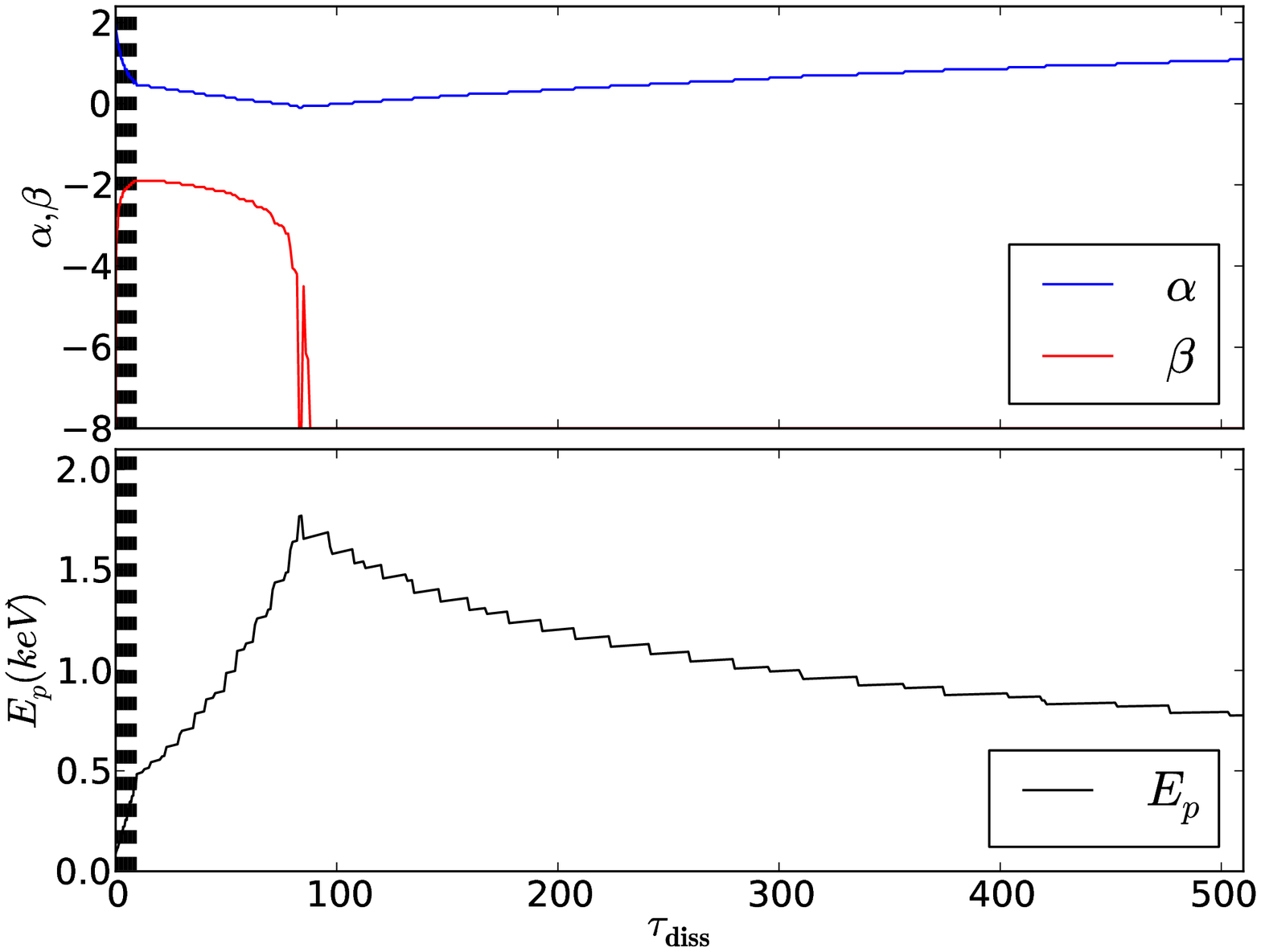}
\vskip 4pt
\caption{Evolution of the Band parameters $\alpha$, $\beta$ and $E_{p}$ of
  spectra from the simulation shown in Figure~\ref{fig:INJ_MB_6_3_W_6_6}.
  The x-axis displays the opacity at which energy deposition occurred.}
\label{fig:ABEPPLOT_INJ_MB_8_3_W_6_6}
\end{figure}
\begin{figure}
\includegraphics[width=80mm]{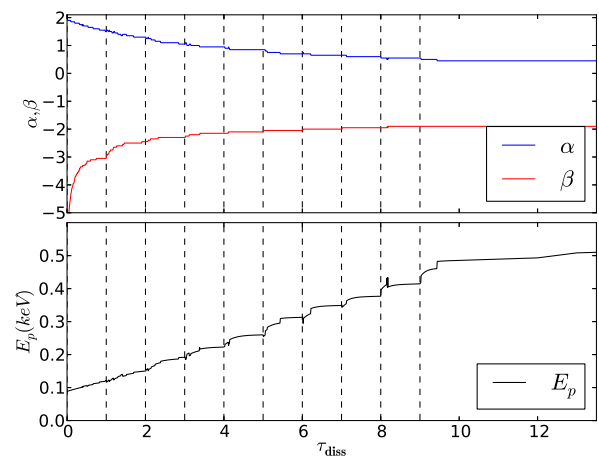}
\vskip 4pt
\caption{Magnified version of
  Figure~\ref{fig:ABEPPLOT_INJ_MB_8_3_W_6_6} depicting the response of
  the Band function parameters $\alpha$, $\beta$ and $E_{p}$ to
  discrete and multiple energy injections, indicated by the broken
  black vertical lines. The x-axis displays the opacity at which
  energy deposition occurred.}
\label{fig:ABEPPLOTCORR_INJ_MB_6_3_W_6_6}
\end{figure}
The results of the MC simulation are shown in
Figure~\ref{fig:INJ_MB_6_3_W_6_6}, where the evolving radiation
spectrum and the spectrum of the kinetic energy of the leptons'
population are displayed. In comparison to
Figures~\ref{fig:MJ_6pt5e9_3_W_6_6} and~\ref{fig:MJNT_8-2.2_3_W_6_6}, 
two differences are apparent for
small optical depths. First, the high-frequency tail develops much
more slowly. Secondly, the slowly developing tail does not extend to
the same high energies and in fact, it never approaches the MeV
mark. Neither of these differences is surprising, given that
a smaller amount of energy is injected at regular intervals. The 
results of the Band function fitting are reported in 
Figure~\ref{fig:ABEPPLOT_INJ_MB_8_3_W_6_6} and bring to our 
attention that like previous other simulations, the
high-frequency photon index $\beta$ is the first to respond, and  
also the first to drop just when the $\alpha$ parameter reaches it's
minimum value. Another remarkable aspect of the multiple injection
scenario is the immediate reaction of the spectrum to new injections,
especially for the high-frequency photon index and the peak frequency
(see Figure~\ref{fig:ABEPPLOTCORR_INJ_MB_6_3_W_6_6}).\\*
What is perhaps mostly interesting, rather than the subtle differences
among the three scenarios discussed here, is the fact the Band
parameters of Figures~\ref{fig:ABEPLOT_MJ_6pt5e9_3_W_6_6},
~\ref{fig:ABEPLOT_MJNT_8-2pt2_3_W_6_6},
and~\ref{fig:ABEPPLOT_INJ_MB_8_3_W_6_6} show remarkably similar
behavior, even though the injection scenarios are very different. In
all three cases, injection at low optical depth only produces a
high-frequency power-law tail. Injection at moderate optical depths
($\tau_{diss}\sim10-100$) produces a high-frequency power-law
tail, a shift in the peak frequency, and a non-thermal low-frequency
tail. Injection at high to very high optical depths only results in a
non-thermal low-frequency tail (see also Section~4 for a discussion).

\subsection{Pair Enriched Plasmas}
In this section we investigate plasmas enriched by $e^-e^+$ pairs, by 
choosing $n_\gamma/n_{\rm{lep}}=10$. GRB plasmas can become pair 
enriched via energy injection through shocks/magnetic dissipation 
(Rees \& Meszaros 2005; Meszaros et al. 2002; Pe'er \& Waxman 2004) 
and if the peak energy of the resulting distribution exceeds 20 keV 
(Svensson 1982). The generation of pairs is also evident from the 
photon and lepton distributions crossing the 511 keV mark as shown in 
the simulations in Sections~\ref{sect:thermal_phrich} 
and~\ref{sect:nonthermal_phrich}. We assume, as in the 
previous scenario, that the the pair 
enriched leptons are impulsively heated by injecting  
equal amounts of kinetic energy $K/10$ 
for the first two simulations, albeit with 
different distribution functions (Maxwellian and Maxwellian plus 
power law). The third simulation explores the spectral evolution of 
a pair enriched plasma with an even greater kinetic energy 
injection. The initial photon count of the plasma  
$N_{\gamma}$ is $1.01\times 10^5$. Being pair-enriched, the 
total lepton count $N_{e}$ of the plasma is $1.01\times 10^4$,
\be
N_{e} = N_{e^-} + 2N_{e^+ e^-} = 10^2 + 10^4
\ee
where $N_{e^-}$ are electrons associated with protons and $N_{e^+
  e^-}$ denotes the number of pairs in the system.
\subsubsection{Maxwellian leptons}
\label{sect:thermal_pairrich}
\begin{figure}
\includegraphics[width=80mm]{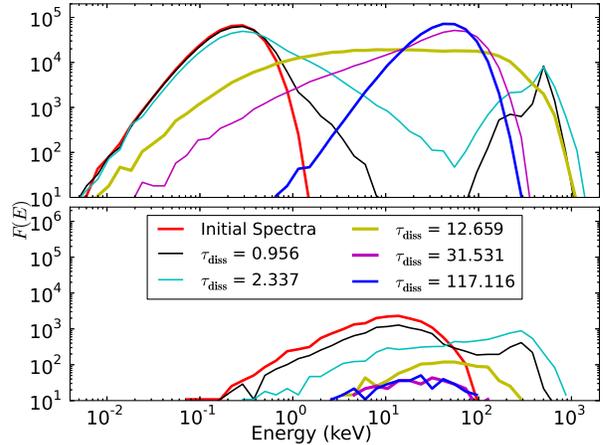}
\vskip 4pt
\caption{Photon spectrum (upper panel) and leptons' kinetic energy
  distribution (lower panel) at different stages of the pair-enriched 
  simulation discussed in Section~\ref{sect:thermal_pairrich}. The legend displays the various 
  optical depths at which if energy was injected, the corresponding color
  coded spectrum and distribution would be observed.}
\label{fig:MB_8_1.01e4_W_6_1.01e5_pairen}
\end{figure}

\begin{figure}
\includegraphics[width=80mm]{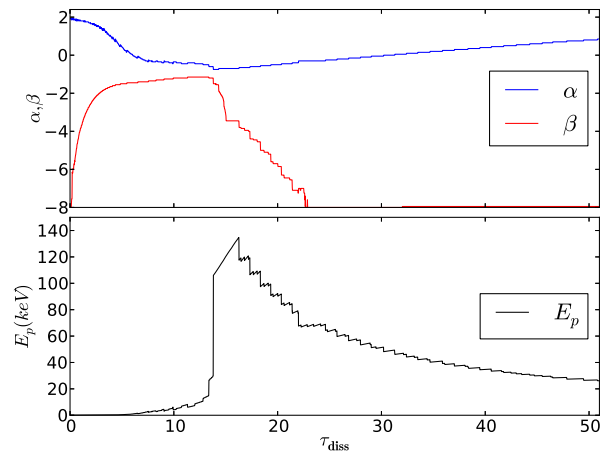}
\vskip 4pt
\caption{Evolution of the Band parameters $\alpha$, $\beta$ and
  $E_{p}$ of spectra from the simulation shown in
  Figure~\ref{fig:MB_8_1.01e4_W_6_1.01e5_pairen} for increasing values
  of energy-injection optical depths.}
\label{fig:ABEPPLOT_MB_8_1.01e4_W_6_1.01e5_pairen}
\end{figure}


We initiate the simulation with Maxwellian pair-enriched leptons that
have been impulsively heated to $10^8$ K, thereby taking the
population out of equilibrium with the photons.  The results of the
simulation are displayed in
Figure~\ref{fig:MB_8_1.01e4_W_6_1.01e5_pairen} with the upper panel
depicting the photon spectra and the lower panel illustrating the
kinetic energies of the leptons. Firstly, as observed in the section
on photon rich plasmas, the final (blue curve) spectra is consistent 
with the equilibrium Wien distribution.  For $\tau_{diss} \sim 1$ a	
bump is observed to spike near the annihilation line along with a  
power law tail (black curve). The lepton distribution also displays a
two component distribution (black curve in the lower panel). For
$\tau_{diss} \sim 2.3$, the power law tail extends farther to high
frequencies and merges with the annihilation bump (cyan curve). On
increasing the injection opacity to around~$13$, the low frequency 
spectrum flattens, the peak frequency increases and the annihilation
bump merges completely with the initial Wien distribution (or the
remnant of the initial spectrum) creating a non-thermal flattened
plateau-like feature (yellow curve). The high-frequency power law tail
returns to the equilibrium Wien spectrum much earlier ($\tau_{diss} \leq
32$) than the non-thermal low frequency tail, which requires about
($\tau_{diss} \sim 100$) to form the equilibrium spectrum. We
interpret this behavior to the inability of the plasma to support 
a large population of pairs. As a
consequence the pairs quickly annihilate and a large amount of
$\sim511$~keV photons are injected in the plasma.\\*
The Band parameters obtained by fitting the Band functions to the 
simulation spectra are plotted
in Figure~\ref{fig:ABEPPLOT_MB_8_1.01e4_W_6_1.01e5_pairen}. We note
that for moderate optical depths, $\alpha=-0.75$ and
$\beta=-1.15$ which corresponds to an extremely non-thermal
spectrum. We also observe from the lower panel of
Figure~\ref{fig:ABEPPLOT_MB_8_1.01e4_W_6_1.01e5_pairen} that
$E_{p}=20-40$~keV. Furthermore, an anti-correlation is
observed between the Band parameters $\alpha$ and $\beta$ and 
between $\alpha$ and $E_p$.
\subsubsection{Maxwellian leptons at $10^8$ K with a power law tail}
\label{sect:MBNT_pairrich}
\begin{figure}
\includegraphics[width=80mm]{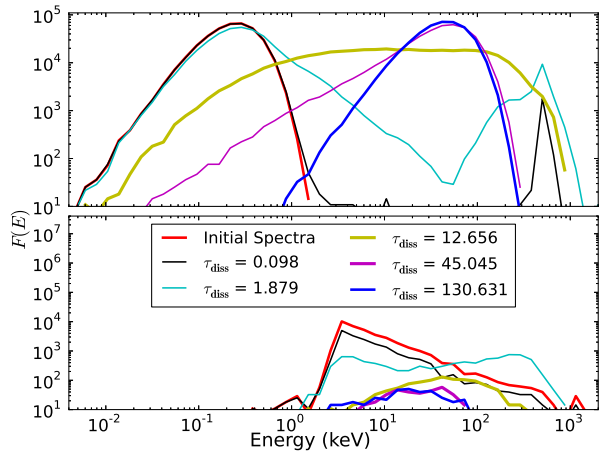}
\vskip 4pt
\caption{Photon spectrum (upper panel) and leptons' kinetic energy
  distribution (lower panel) at different stages of the simulation 
  discussed in Section~\ref{sect:MBNT_pairrich}. The legend associates the various 
  optical depths of energy injection with the 
  corresponding color coded spectrum and distribution observed.}
\label{fig:MBNT_8-2.2_1.01e4_W_6_1.01e5_pe}
\end{figure}
\begin{figure}
\includegraphics[width=80mm]{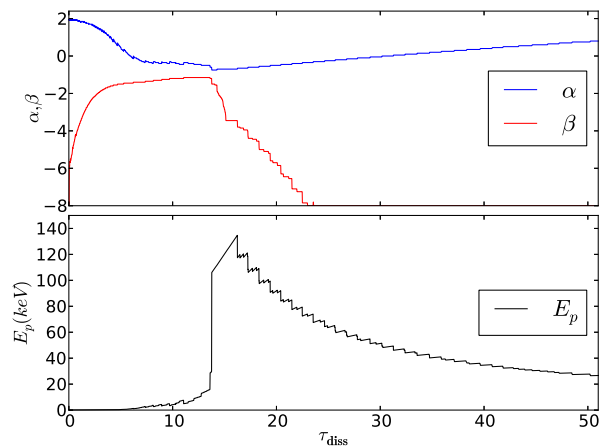}
\vskip 4pt
\caption{Evolution of the Band parameters $\alpha$, $\beta$ and $E_{p}$ of
  spectra from the simulation shown in Figure~\ref{fig:MBNT_8-2.2_1.01e4_W_6_1.01e5_pe}.
  The x-axis displays the opacity at which energy injection occurred.}
\label{fig:ABEPPLOT_MBNT_8-2.2_1.01e4_W_6_1.01e5}
\end{figure}
This simulation initializes the lower energy lepton population 
as thermally distributed at $10^8$~K and a higher energy 
population with a power law tail. However, pair enrichment 
and constraining the injected kinetic energy to $K/10$ lowers  
the average kinetic energy per lepton in comparison to the 
photon-rich plasmas. As a result the leptons are generated 
according to the distribution 
\be
N(E)dE \propto (\gamma - 1)^{-p} d(\gamma - 1)
\ee
where $\gamma$ is the lepton's Lorentz factor and $p=2.2$.
The red curve in lower panel of 
Figure~\ref{fig:MBNT_8-2.2_1.01e4_W_6_1.01e5_pe} displays the initial
kinetic energy distribution of the lepton population. Note that the 
power law tail does not extend to high energies as the tail in 
Figure~\ref{fig:MJNT_8-2.2_3_W_6_6} does. 
The figure also shows the evolution of the photon spectra and leptons' 
kinetic energy as equilibrium restoration occurs. For the photons, the 
initial spectra (red curve) and equilibrium spectrum (blue curve) fit 
the Wien distribution. As is expected, pair annihilation produces a 
hump in the vicinity of the 511 keV region. Meanwhile, the photons 
forming the initial Wien spectrum 
form a power law tail. Similar to the previous scenario, at around
$\tau_{diss} \sim 2$ , the power law extends to high frequencies and
merges with the growing annihilation hump (cyan curve). We also 
observe a two component distribution in the lepton panel. By
$\tau_{diss} \sim 13$, the two component spectrum transforms into a
broad band flat-plateau like spectrum (yellow curve) with the
low-frequency spectrum being modified as well. The high frequency
spectrum of the magenta curve (for $\tau_{diss} \sim 45$) assumes 
the exponential cut-off of the Wien spectrum while 
the low-frequency tail is still prominent. These features 
make the transient spectra highly non-thermal.\\*
A comparison of Figure~\ref{fig:MB_8_1.01e4_W_6_1.01e5_pairen} with 
Figure~\ref{fig:MBNT_8-2.2_1.01e4_W_6_1.01e5_pe} informs us that 
the spectra of these two scenarios are quite similar. Consequently, a 
comparison among 
Figure~\ref{fig:ABEPPLOT_MB_8_1.01e4_W_6_1.01e5_pairen} and 
Figure~\ref{fig:ABEPPLOT_MBNT_8-2.2_1.01e4_W_6_1.01e5} also exhibits 
very similar results - including the anti-correlations 
between $\alpha$ and peak frequency and between $\alpha$ and $\beta$.

\subsubsection{Maxwellian leptons at $10^8$ K with a power law tail $p=2.2$}
\label{sect:nonthermal_pairrich}
This section explores the system when pair enriched leptons 
are distributed according the Maxwell-Boltzmann distribution at 
$10^8$~K for lower energies whereas the high energy ones form 
a power-law tail with index $p=2.2$.

\begin{figure}
\includegraphics[width=80mm]{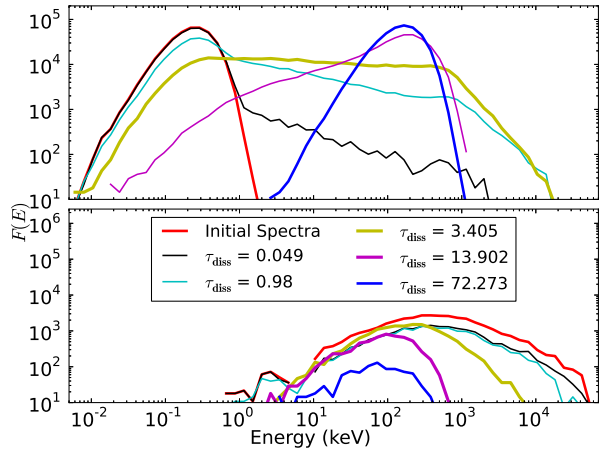}
\vskip 4pt
\caption{Photon spectrum (upper panel) and leptons' kinetic energy
  distribution (lower panel) at different stages of the pair-enriched 
  simulation discussed in Section~\ref{sect:nonthermal_pairrich}.
  The legend associates the various 
  optical depths of energy injection with the 
  corresponding color coded particle spectrum and distribution.}
\label{fig:MJNT_8-2.2_1.01e4_W_6_1.01e5_pe_elpos}
\end{figure}
\begin{figure}
\includegraphics[width=80mm]{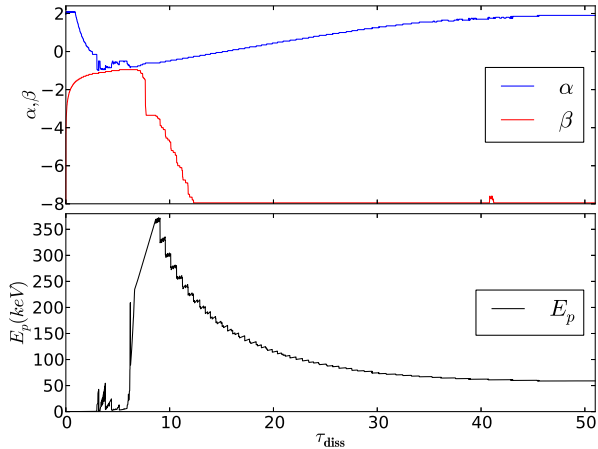}
\vskip 4pt
\caption{Evolution of the Band parameters $\alpha$, $\beta$ and $E_{p}$ of
  spectra from the simulation shown in 
  Figure~\ref{fig:MJNT_8-2.2_1.01e4_W_6_1.01e5_pe_elpos} with  
  increasing energy-injection opacity.}
\label{fig:ABEPPLOT_MJNT_8-2.2_1.01e4_W_6_1.01e5}
\end{figure}
\begin{figure}
\includegraphics[width=80mm]{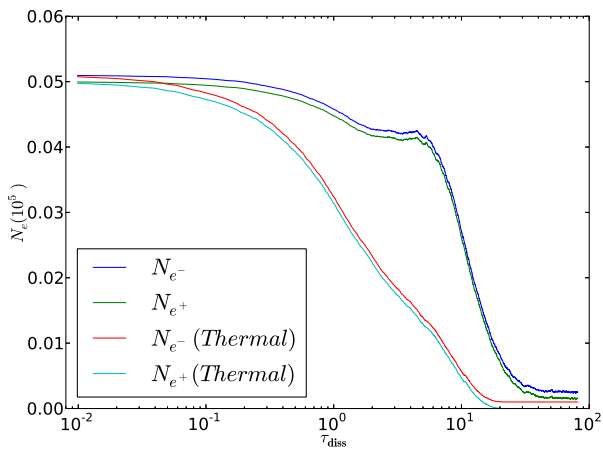}
\vskip 4pt
\caption{Evolution of lepton count $N_{e}$ for the simulations 
in Section~\ref{sect:thermal_pairrich} (labeled as $Thermal$) 
and~\ref{sect:nonthermal_pairrich}. Note that the lepton count 
evolution of the simulations discussed in 
Sections~\ref{sect:thermal_pairrich} and~\ref{sect:MBNT_pairrich} 
is indistinguishable. The x-axis displays 
the opacity at which energy injection occurred.}
\label{fig:NLEP_MJNT_8-2.2_1.01e4_W_6_1.01e5}
\end{figure}
Similar to the previously discussed cases, the photon spectrum fits
the Wien spectrum at equilibrium in
Figure~\ref{fig:MJNT_8-2.2_1.01e4_W_6_1.01e5_pe_elpos}. A remarkable
difference between
Figure~\ref{fig:MJNT_8-2.2_1.01e4_W_6_1.01e5_pe_elpos}, and between
Figures~\ref{fig:MB_8_1.01e4_W_6_1.01e5_pairen}
and~\ref{fig:MBNT_8-2.2_1.01e4_W_6_1.01e5_pe} is that the high
frequency power law tail catches up with the pair-annihilation much
earlier $(\tau_{diss} << 0.05)$ as depicted by the black curve.
Remnants of the hump are visible in the black and cyan curves.
Furthermore, for less than 1 scatterings, the low frequency tail
becomes softer than the Wien spectrum (cyan curve). Another important
non-thermal feature is the broadband nature of the flattened spectrum
(the yellow curve extends over four orders of magnitude in
frequency). By about $\tau_{diss} \sim 14$, the truncated high
frequency tail approaches the exponential cut-off of the Wien
spectrum, whereas the soft low
frequency tail still persists.\\*
The best-fit Band function obtained by $\chi^2$ minimization
technique, produces highly non-thermal spectral indices
  ($\alpha$ and $\beta$) but the peak frequency as shown in
  Figure~\ref{fig:ABEPPLOT_MJNT_8-2.2_1.01e4_W_6_1.01e5} is relatively
  high for GRBs. The lack of smoothness in the $\alpha$ values for
moderate optical depths is due to the flatness of the photon spectrum
as seen from the yellow curve in
Figure~\ref{fig:MJNT_8-2.2_1.01e4_W_6_1.01e5_pe_elpos}, which occurs
in conjunction with the transient saturation phase in the lepton count
(see Figure~\ref{fig:NLEP_MJNT_8-2.2_1.01e4_W_6_1.01e5}).
Figure~\ref{fig:NLEP_MJNT_8-2.2_1.01e4_W_6_1.01e5} also displays and
compares the lepton count for the simulation in
Section~\ref{sect:thermal_pairrich} (the curves labeled as $Thermal$,
which are indistinguishable from the pair evolution in
Section~\ref{sect:MBNT_pairrich}). Although the initial lepton content
of the plasmas in the three discussed simulations is identical, the
plasma with a greater kinetic energy injection can sustain pairs for
larger optical depths leading to a much broader and flatter spectrum.
For moderate number of scatterings, we obtain $\alpha=-1$ and
$\beta=-0.95$. Again, an anti-correlation is found to exist between
the parameters $\alpha$ and $\beta$ and also between $\alpha$ and the
peak frequency.\\*
\section{Summary and Discussion}
We present Monte Carlo simulations of Compton scattering, $e^-e^+$
pair production, and $e^-e^+$ pair annihilation in GRB 
fireballs subject to mild to moderate internal dissipation. We explore
cases of photon-rich media -- as expected in baryonic fireballs -- and
of pair-dominated media.  The leptonic component in our simulations is
initially set out of equilibrium by a sudden injection of energy and
the spectrum is followed as continuous collisions among photons and
leptons restore equilibrium.
\begin{figure}
\includegraphics[width=\columnwidth]{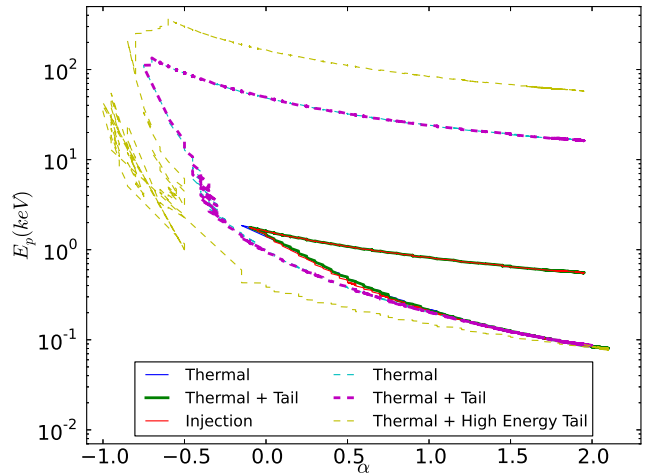}
\vskip 4pt
\caption{Plot of Band parameters $E_p$ and $\alpha$ for the various 
  simulations discussed. The solid curves represent photon-rich 
  plasmas ($\frac{N_{\gamma}}{N_{lep}} = 1000$) whereas the broken 
  curves are indicative of pair-enriched plasmas where 
  ($\frac{N_{\gamma}}{N_{lep}} = 10$). Note the similarity among 
  the curves and the exhibited anti-correlation.}
\label{fig:alpha_ep}
\end{figure}
\begin{figure}
\includegraphics[width=\columnwidth]{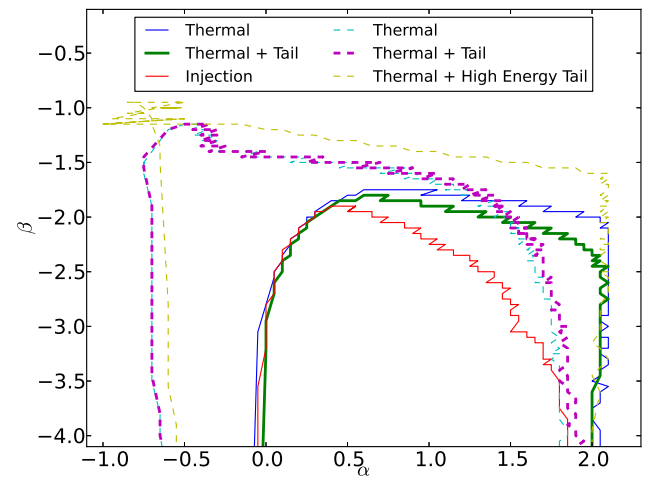}
\vskip 4pt
\caption{Plot of Band parameters $\beta$ and $\alpha$ for the various 
  simulations discussed. The solid curves represent photon-rich 
  plasmas ($\frac{N_{\gamma}}{N_{lep}} = 1000$) whereas the broken 
  curves are indicative of pair-enriched plasmas where 
  ($\frac{N_{\gamma}}{N_{lep}} = 10$). Note the complex behavior of 
  the curves, especially the evolution of $\beta$.}
\label{fig:alpha_beta}
\end{figure}

We find that non-thermal spectra arise from transient effects. Such
spectra could be advected by the expanding fireball and released
before equilibrium is reached if the dissipation takes place at
optical depths of up to several hundred.  We show that the transient
spectra can be reasonably fit by a Band function (Band et al. 1993)
within a frequency range of 2-3 orders of magnitude around the
  peak and could therefore explain GRB observations. As suggested by
Lazzati \& Begelman (2010), non-thermal features can arise even if
both the photon and lepton distributions are initially thermal,
provided that they are at different temperatures. As a matter of fact,
we find that the spectrum emerging from the fireball after a
dissipation event at a certain optical depth does not depend strongly
on the way in which the energy was deposited in the leptons. For the
photon-rich cases, the first reaction of the photon spectrum to a
sudden energy injection into the leptons is the formation of a
high-frequency power-law, either because non-thermal leptons are
present or through the mechanism described in Lazzati \& Begelman
(2010). If the injection happens at moderate optical depths, the peak
frequency of the photon spectrum also shifts to higher frequencies and
a non-thermal low-frequency tail appears.  If the energy injection
occurs at somewhat large optical depths, the high frequency tail
disappears and the spectrum presents a cutoff just above the peak. The
low-frequency non-thermal tail is however very resilient and only if
the dissipation takes place at very large optical depths, the
equilibrium Wien spectrum is attained. The pair-enriched simulations
show a more complex behavior at low optical depths due to pair
processes, however we still observe the low frequency tail's resilient
behavior. We show that this phenomenology is rather independent on the
details of the energy dissipation process and generated lepton
distributions: non-thermal leptons, high-temperature thermal leptons,
and multiple discrete injection events all produce similar
spectra. For the case of the pair-enriched simulations however, we
obtain peak frequencies that are somewhat large in the comoving frame
(several hundred keV) making this scenario less interesting for
explaining observed burst spectra. However their complex behavior and
extreme peak energies offer a tantalizing explanation for the rich
diversity observed in peak energies of GRBs (Goldstein et al 2012)
especially when the peak energies $>$ MeV.

The conclusion we can glean from this study is therefore that
comptonization of advected seed photons by sub-photospheric
dissipation continues to be a viable model to explain the prompt
gamma-ray bursts spectrum. Agreement is particularly strong when the
dissipation occurs at moderate optical depth (of the order of tens) so
that both a high- and a low-frequency tail are produced.  Dissipation
at too low optical depth would only produce a high-frequency tail,
while dissipation at too large optical depth would only produce a
low-frequency tail. In a GRB dissipation is likely to occur at all
optical depths (e.g. Lazzati et al. 2009). The dissipation events that
occur at moderate optical depth would therefore be those mostly
affecting the spectrum and giving it its non-thermal appearance. 
Bursts characterized by a Band spectrum over more than three orders 
of magnitude of frequency remain however challenging for this model, 
and other effects need to be invoked to avoid deviations from the 
pure power-law behavior at very low and high frequencies. Among 
these effects, some studied in the literature are sub-photospheric, 
radiation mediated multiple shocks (Keren \& Levinson 2014), line of  
sight effects (Pe'er \& Ryde 2011) and high-latitude emissions (Deng 
\& Zhang 2014).

\subsection{Spectral correlations}
Besides finding that the overall shape of the partially Comptonized
spectra is qualitatively analogous to what observed in GRBs, we find
that this model predicts the existence of two correlations that can
be used as a test of its validity. We first notice an anti-correlation
between the low-energy photon index $\alpha$ and the peak frequency.
The correlation is clearly seen in Figure~\ref{fig:alpha_ep}, where
results from all simulations are shown simultaneously.  All
simulations start with the same injected photon spectrum, the common
point in the lower right of the diagram. The leptons in all three of
the photon-rich simulations are energized to identical total kinetic
energies $K$ albeit different distribution functions. It is clear that
the evolution of all photon-rich simulations is virtually
indistinguishable from each other. As more and more scatterings
occur, the peak frequency initially grows and the low-frequency slope
flattens.  At moderate optical depths ($\sim100$ in all three cases)
the peak frequency reaches its maximum, the high frequency tail
disappears (shown in the Figure~\ref{fig:alpha_beta}) and the 
low-frequency tail begins to thermalize, dragging the peak frequency to slightly lower
values. The correlation has two branches, a steeper one for $\tau<100$
and a flatter one at $\tau>100$. The second branch corresponds,
however, to spectra without a high-frequency tail and is therefore not
expected to represent observed GRBs. A similar pattern is followed by
the pair-enriched cases, with the main difference that larger peak
frequencies are attained along with softer values for $\alpha$ and
$\beta$. The evolutionary curves for the pair-enriched cases show
complexity due to the presence of pairs especially at low opacities -
with the simulation in Section~\ref{sect:nonthermal_pairrich} showing a 
greater amount of variability due to its ability to sustain pairs 
by temporarily balancing the
number of pair production and annihilation events (see
Figure~\ref{fig:NLEP_MJNT_8-2.2_1.01e4_W_6_1.01e5}).\\*
In addition to the $\alpha-\nu_{\rm{pk}}$ anti-correlation, we also 
find hints of an anti-correlation between $\alpha$ and $\beta$. This
correlation is shown in Figure~\ref{fig:alpha_beta} and is much more
complex, reflecting the more complex behavior of the high-frequency
spectrum with respect to the low-frequency one. In the case of the
high-frequency photon-index $\beta$, the way in which the energy is
injected in the lepton population matters, each simulation producing a
different track on the graph.\\*

Comparing these predictions to GRB spectral data is not
straightforward, since the correlations should not be strong in
observational data. Adding together data from different bursts, the
correlations in the observer frame would be diluted by the different
bulk Lorentz factors of bursts and by the diversity of the particle
ratio, radiation temperature, and dissipation intensity among busts
and pulses in a single burst. Still, some degree of correlation has
been discussed in the literature, with contradictory conclusions as to
its robustness. The $\alpha-\nu_{\rm{pk}}$ anti-correlation has been
discussed in large burst samples (e.g. Amati et al. 2002; Goldstein at
al. 2012; Burgess et al. 2014). The $\alpha$-$\beta$ anti-correlation 
has been observed for some bursts (Zhang et al. 2011), however it is 
not a common feature among GRBs. 

Photospheric dissipation models have found it difficult to reproduce
low frequency photon index $\alpha \sim -1$ and have been unable to
explain the GeV emissions (Zhang et al 2011).
Figure~\ref{fig:MJNT_8-2.2_1.01e4_W_6_1.01e5_pe_elpos} displays the
emission spectra in the rest frame of the burst and once Lorentz
boosted the photons forming the high frequency tail reach GeV
energies. For low/moderate opacities, our simulations have
consistently reproduced the low-energy photon index $\alpha<0$ as
shown in Figure~\ref{fig:alpha_ep} thus providing a possible
resolution for the mentioned issues.  Our current model is unable to
reproduce $\alpha < -1.1$ for the parameter space explored, however
additional effects such can modify and
further soften the low frequency spectra.  Analogous studies of
comptonization effects in GRB outflows have been performed in the
past, for example by Giannios (2006) and Pe'er et al. (2006). Our work
differs from both of these previous studies in both content and
methodology. Giannios (2006) studied with Monte Carlo techniques the
formation of the spectrum in magnetized outflows, considering a
particular form of dissipation and assuming that the electrons
distribution is always thermalized, albeit at an evolving
temperature. Pe'er et al. (2006), instead, used a code that solves the
kinetic equations for particles and photons, and considered injection
of non-thermal particles (as in our
Section~\ref{sect:nonthermal_phrich}) as well as continuous injection
of energy in a thermal distribution. None of these previous studies
consider impulsive injection of energy in thermal leptons, as
discussed here or the case of multiple, discrete injection events. In
an attempt to keep our results as general as possible we have
performed the calculations in a static medium, rather than in an
expanding jet. As long as the opacity at which the dissipation occurs
is not too large, this should not be a major limitation, and the
advantage is that our results are not limited to a particular
prescription for the jet radial evolution. In addition, most of the
interesting results (the non-thermal spectra) are obtained for small
and moderate values of the optical depth (or, analogously, of the
number of scatterings that take place before the radiation is
released). It should also be noted that the assumption of an impulsive
acceleration of the leptons that does not affect the photon spectra is
likely not adequate in a highly opaque medium. A final limitation of
this study is that only moderate values of the particle ratio can be
explored. This is an inevitable limitation when both the lepton and
photon distributions are followed in the scattering process with a
Monte Carlo technique. If one of the two significantly outnumbers the
other, a very large number of photons (or leptons) are required,
making the calculation extremely challenging and would require
parallelizing the code. While performing such simulations is important
and will eventually become possible, we do not anticipate big
phenomenological differences with respect to what we consider
here. Even with less electrons, we expect the formation of a
high-frequency tail (e.g. Lazzati \& Begelman 2010), the subsequent
shift of the peak frequency accompanied by a flattening of the
low-frequency photon index, and complete thermalization only after
many scatterings (i.e., only if the dissipation occurs at a very high
optical depth).

\section*{Acknowledgments}
We thank the anonymous referee for her/his comments leading to 
improvement and clarity of the manuscript. We thank Paolo Coppi for
his advice and insight into the physics of scattering and Gabriele
Ghisellini and Dimitrios Giannios for insightful discussions. This
work was supported in part by NASA Fermi GI grant NNX12AO74G and NASA
Swift GI grant NNX13AO95G.

\label{lastpage}


\begin{thebibliography}{99}
\bibitem[(]{} Amati, L., Frontera, F., Tavani, M., et al. 2002, A\&A, 390, 81
\bibitem[(]{} Amati L. 2006, MNRAS, 372, 233
\bibitem[(]{} Asano, K., Inoue, S., \& M{\'e}sz{\'a}ros P. 2010, \apj, 725, L121 
\bibitem[(]{} Band, D., Matteson, J., Ford, L., et al. 1993, \apj, 413, 281
\bibitem[(]{} Blumenthal, G.R., \& Gould, R.J. 1970, Rev. Mod. Phys, 42, 237
\bibitem[(]{} Beloborodov A.~M. 2010, MNRAS, 407, 1033
\bibitem[(]{} Beloborodov A.~M. 2013, ApJ, 764, 157
\bibitem[(]{} Bo{\v s}njak {\v Z}., Daigne F., \& Dubus G. 2009, A\&A, 498, 677
\bibitem[(]{} Burgess, J. M., Ryde, F., \& Yu, H.-F. 2014, arXiv:1410.7647
\bibitem[(]{} Coppi P.S., \& Blandford R.D. 1990, MNRAS, 245, 453
\bibitem[(]{} Crumley P., \& Kumar P. 2013, MNRAS, 429, 3238
\bibitem[(]{} Daigne F., Bo{\v s}njak {\v Z}., \& Dubus G. 2011, A\&A, 526, A110
\bibitem[(]{} Deng, W., \& Zhang, B. 2014, \apj, 785, 112
\bibitem[(]{} Fan Y.-Z., Wei D.-M., Zhang F.-W., et al. 2012, ApJ, 755, L6
\bibitem[(]{} Ghirlanda G., Nava L., Ghisellini G., et al. 2012, MNRAS, 420, 483
\bibitem[(]{} Ghisellini G., Celotti A., \& Lazzati D. 2000, MNRAS, 313, L1
\bibitem[(]{} Ghisellini G. 2010, AIPC, 1248, 45f
\bibitem[(]{} Giannios D. 2006, A\&A, 457, 763
\bibitem[(]{} Giannios D., \& Spruit H.~C. 2006, A\&A, 450, 887
\bibitem[(]{} Goldstein, A., Burgess, J.M., Preece, R.D., et al. 2012, \apjs, 199, 19
\bibitem[(]{} Gould, R.J., \& Schreder, G.P. 1967, Phy.Rev., 155, 1404
\bibitem[(]{} Guetta, D., Spada, M., \& Waxman, E. 2001, \apj, 557, 399
\bibitem[(]{} Guiriec, S., Connaughton, V., Briggs, M., et al. 2011, \apj, 727, 33
\bibitem[(]{} Guiriec, S., Daigne, F., Hasco{\"e}t, R., et al. 2013, \apj, 770, 32
\bibitem[(]{} Hasco{\"e}t R., Daigne F., \& Mochkovitch R. 2013, A\&A, 551,
  A124
\bibitem[(]{} Ito,H., Nagataki, S., Ono, M., et al. 2013, ApJ, 777, 62
\bibitem[(]{} Ito H., Nagataki S., Matsumoto J., et al. 2014, arXiv:1405.6284
\bibitem[(]{} Jauch, J.M., \& Rohrlich, F. 1980, Theory of Photons and Electrons (2nd Extended ed.; New York: Springer-Verlag.)
\bibitem[(]{} Keren, S., \& Levinson, A. 2014, \apj, 789, 128 
\bibitem[(]{} Lazzati D., Morsony B.~J., \& Begelman M.~C. 2009, ApJ, 700,
  L47
\bibitem[(]{} Lazzati D., \& Begelman M.~C. 2010, ApJ, 725, 1137
\bibitem[(]{} Lazzati D., Morsony B.~J., Margutti R., et al. 2013, ApJ, 765, 103
\bibitem[(]{} Liang E.-W., Yi S.-X., Zhang J., et al. 2010, 
ApJ, 725, 2209
\bibitem[(]{} Lloyd N.~M., \& Petrosian V. 2000, ApJ, 543, 722
\bibitem[(]{} Longair, Malcolm, S. 2011, High Energy
    Astrophysics, (New York: Cambridge)
\bibitem[(]{} L{\'o}pez-C{\'a}mara D., Morsony B.~J., Begelman M.~C., et al. 2013, ApJ, 767, 19
\bibitem[(]{} L{\'o}pez-C{\'a}mara D., Morsony B.~J. \& Lazzati D. 2014,
  MNRAS, 442, 2202
\bibitem[(]{} Lundman C., Pe'er A., \& Ryde F. 2013, MNRAS, 428, 2430
\bibitem[(]{} Massaro F., \& Grindlay J.~E. 2011, ApJ, 727, L1
\bibitem[(]{} Mastichiadis A., \& Kazanas D. 2009, ApJ, 694, L54
\bibitem[(]{} McKinney J.~C. \& Uzdensky D.~A. 2012, MNRAS, 419, 573
\bibitem[(]{} Medvedev M.~V., Pothapragada S.~S. \& Reynolds S.~J. 2009, 
ApJ, 702, L91
\bibitem[(]{} M{\'e}sz{\'a}ros, P., Ramirez-Ruiz, E., Rees, M. J., et al. 2002, ApJ, 578, 812
\bibitem[(]{} M{\'e}sz{\'a}ros P., \& Rees M.~J. 2000, ApJ, 530, 292
\bibitem[(]{} Mizuta A., Nagataki S., \& Aoi J. 2011, ApJ, 732, 26
\bibitem[(]{} Nagakura H., Ito H., Kiuchi K., et al. 2011, ApJ, 731,
  80
\bibitem[(]{} Pe'er A., M{\'e}sz{\'a}ros P., \& Rees M.~J. 2005, ApJ, 635, 476
\bibitem[(]{} Pe'er A., M{\'e}sz{\'a}ros P., \& Rees M.~J. 2006, \apj, 642, 995
\bibitem[(]{} Pe'er A., \& Ryde, F., 2011, \apj, 732, 49
\bibitem[(]{} Pe'er, A., \& Waxman, E. 2004, \apj, 613, 448
\bibitem[(]{} Piran T., 1999, PhR, 314, 575
\bibitem[(]{} Preece R.~D., Briggs M.~S., Mallozzi R.~S., et al. 1998, \apj, 506, L23
\bibitem[(]{} Rees M.~J., \& Meszaros P. 1994, ApJ, 430, L93
\bibitem[(]{} Rees M.~J., \& M{\'e}sz{\'a}ros P. 2005, ApJ, 628, 847
\bibitem[(]{} Resmi L., \& Zhang B. 2012, MNRAS, 426, 1385
\bibitem[(]{} Rybicki, G.B. \& Lightman, A.P. 1979, Radiative
    Processes in Astrophysics, (New York: John Wiley)
\bibitem[(]{} Ryde, F., \& Pe'er A. 2009, ApJ, 702, 1211
\bibitem[(]{} Svensson, R. 1982, \apj, 258, 335
\bibitem[(]{} Zhang, B., \& Yan H. 2011, \apj, 726, 90
\bibitem[(]{} Zhang, B.B., Zhang, B., Liang, E.-W., et al. 2011, \apj, 730, 141
\end{thebibliography}
\end{document}